# Bottle microresonator broadband and low repetition rate frequency comb generator

V. DVOYRIN [1] AND M. SUMETSKY[1,*]

[1]Aston Institute of Photonic Technologies, Aston University, Birmingham B4 7ET, UK
*Corresponding author: m.sumetsky@aston.ac.uk

**We propose a new type of broadband and low repetition rate frequency comb generator which has the shape of an elongated and nanoscale-shallow optical bottle microresonator created at the surface of an optical fiber. The free spectral range (FSR) of the broadband azimuthal eigenfrequency series of this resonator is the exact multiple of the FSR of the dense and narrowband axial series. The effective radius variation of the microresonator is close to a parabola with a nanoscale height which is greater or equal to $\lambda/2\pi n_0$ (here $\lambda$ is the characteristic radiation wavelength and $n_0$ is the refractive index of the microresonator material). Overall, the microresonator possesses a broadband, small FSR, and accurately equidistant spectrum convenient for the generation of a broadband and low repetition rate optical frequency comb. It is shown that this comb can be generated by pumping with a cw laser, which radiation frequency matches a single axial eigenfrequency of the microresonator, or, alternatively, by pumping with a mode-locked laser, which generates a narrowband low repetition rate comb matching a series of equidistant axial eigenfrequencies situated between adjacent azimuthal eigenfrequencies.**

An optical frequency comb (OFC) is defined as an optical spectrum consisting of a series of equally spaced resonances. OFCs with dramatically accurate periodicity have been demonstrated using femtosecond mode-locked lasers [1-5] and high Q-factor optical microresonators [6-10]. The interest to the research and development of OFCs is caused by their important current and potential applications in science and technology ranging from precision molecular spectroscopy to optical clock, quantum computing, and optical communications.

While the mode-locked lasers required for the generation of OFCs are large in dimensions, their alternative, optical microresonators, have a small size and, potentially, can be used for applications where the small dimensions and small power consumption is critical. However, in order to generate an OFC of relatively low repetition rate (RR), the size of toroidal microresonators [4-8, 10] has to be increased inverse proportionally to the RR value. For example, while the OFC RR for the silica toroidal microresonator having 150 μm radius is 220 GHz, the radius of such resonator should be increased to 12.5 mm to generate a comb with 2.6 GHz RR, which has been experimentally demonstrated in [10]. In order to arrive at an order of magnitude lower RR, e.g., equal to 100 MHz, the radius of the toroidal resonator should be increased to impractical 325 mm. To the best of the author's knowledge an optical resonator which is small in dimensions and simultaneously generates broadband and low RR combs has not yet been proposed.

In this Letter, we describe a microresonator which potential experimental realization is supposed to solve the problem indicated [11]. The idea of our device is illustrated in Fig. 1. The microresonator under consideration is a bottle resonator [12] created at the surface of an optical fiber (Fig. 1(a)). The resonator is strongly elongated along the fiber axis $z$. It is assumed that the axial variation of the resonator radius $\Delta r(z)$ (more precisely, the effective radius variation (ERV) [13]) is adiabatically smooth and is close to parabolic in a finite vicinity of the microresonator center $z_0$ (Fig. 1(a)). The ERV in this vicinity is small, $|\Delta r(z)| << r_0$, having the order of a hundred nanometers only (see calculations below). For this reason, propagation of whispering gallery modes (WGMs) in this vicinity is similar to those described by the Surface Nanoscale Axial Photonics (SNAP) theory [13]. Generally, due to the adiabatically slow variation of $r(z)$, the WGMs of the bottle resonator considered can be found by adiabatic separation of variables in the cylindrical frame of reference $(\rho, \varphi, z)$. They are described by the azimuthal, axial, and radial quantum numbers, respectively, $m$, $q$, and $p$ [12]:

$$E_{m,p,q}(\rho,\varphi,z) = \exp(im\varphi) \cdot U_{m,p}\left(\frac{\rho}{r(z)}\right) \cdot \Psi_{m,p,q}(z) \quad (1)$$

Here we are interested in the case of large azimuthal and axial quantum numbers, $m, q >> 1$, when the WGM eigenfrequencies are determined by the semiclassical quantization rule [12].

OFCs generated at the eigenfrequencies of WGM resonators determined by their azimuthal quantum number $m$ have been demonstrated and studied both experimentally and theoretically [6-8, 10]. The axial eigenmodes of a bottle resonator were experimentally demonstrated for the first time in [9]. In [14], the numerical simulation of narrowband low RR OFCs generated at a series of axial eigenfrequencies in a shallow bottle resonator was performed. Here we are interested in broadband and low RR OFCs simultaneously generated at the axial and azimuthal eigenfrequencies. For the effective generation of these combs, the equidistant separation of the corresponding axial eigenfrequencies is required. The shape of an adiabatically elongated bottle resonator $r_{equid}(z)$ having equidistant axial eigenfrequencies with separation $\Delta\nu_{ax}$ has been determined in [12]:

$$r_{equid}(z) = r_0 \cos\left(\frac{2\pi n_0}{c} \Delta\nu_{ax}(z-z_0)\right) \quad (2)$$

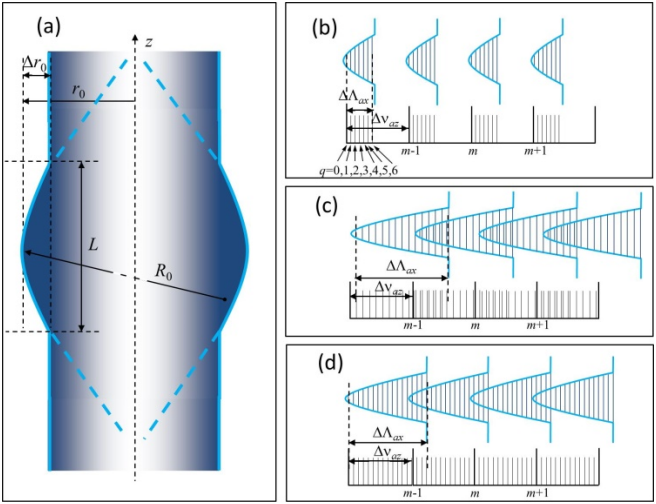

Fig. 1. Illustration of a bottle microresonator enabling the generation of a broadband and low RR OFC and its spectrum. The microresonator has a shape of a shallow and elongated bump $r(z)$ shown in (a) (bold blue solid curve) which coincides with the cosine profile $r_{equid}(z)$ (dashed bold blue curve) defined by Eq. (2) for $r_0 - r(z) < \Delta r_0$, where $\Delta r_0 \ll r_0$ determines the height of the bottle resonator. (b) – Illustration of the spectrum of a bottle microresonator having the full width of axial spectral series $\Delta\Lambda_{ax}$ smaller than the azimuthal FSR $\Delta\nu_{az}$. (c) – Spectrum of a bottle having the full width of axial spectral series $\Delta\Omega_{ax}$ greater than the azimuthal FSR $\Delta\nu_{az}$. (d) – Spectrum of a bottle having the full width of axial spectral series $\Delta\Lambda_{ax}$ greater than the azimuthal FSR $\Delta\Lambda_{az}$ and the azimuthal FSR $\Delta\nu_{az}$ which is an exact multiple of the axial FSR $\Delta\nu_{ax}$.

where $\Delta\nu_{ax}$ is the axial free spectral range (FSR), $n_0$ is the refractive index of the bottle resonator material, and $c$ is the speed of light. For this shape (solid and dashed blue curve in Fig. 1(a)), the semiclassical expression for eigenfrequencies is [12, 15]

$$\nu^\pm_{m,p,q} = \frac{c}{2\pi n_0 r_0}\left[m + \zeta_p\left(\frac{m}{2}\right)^{1/3} - \frac{n_0^{\pm 1}}{(n_0^2-1)^{1/2}}\right] + \left(q + \frac{1}{2}\right)\Delta\nu_{ax} \quad (3)$$

where $\zeta_{1,2,3,...} = 2.338, 4.088, 5.521...$ are the roots of the Airy function, sign $\pm$ determines the mode polarization, and the dependence on quantum numbers $m$ and $p$ is borrowed from [15]. In the vicinity of the microresonator center $z_0$ of our interest, the bottle shape can be approximated by a parabola, $r(z) = r_0 - (z-z_0)^2/(2R_0)$, where $R_0$ is the axial radius at $z = z_0$. From Eq. (2) and (3) we find $\Delta\nu_{ax} = c(2\pi n_0)^{-1}(r_0 R_0)^{-1/2}$. Remarkably, the shape of the bottle determined by Eq. (2) does not depend on quantum numbers, while the dependence on quantum numbers $m, p$ and $q$ are separated in Eq. (3). Thus, the eigenfrequency spacing $\Delta\nu_{ax}$ can be made the same along the whole frequency bandwidth under interest. For example, the azimuthal FSR of a silica bottle resonator with $r_0 = 300$ μm is approximately $\Delta\nu_{az} \simeq c/(2\pi n_0 r_0) \simeq 109$ GHz. A bottle resonator will have the axial FSR $\Delta\nu_{ax} = 1$ GHz if its axial radius is $R_0 = c^2/(4\pi^2 n_0^2 r_0 \Delta\nu_{ax}^2) \simeq 3.4$ m, while a ten times smaller $\Delta\nu_{ax} = 100$ MHz corresponds to $R_0 = 340$ m. Remarkably, shallow bottle resonators with such gigantic axial radii can be fabricated (e.g., a semi-parabolic bottle resonator experimentally demonstrated in [16] had $R_0 = 1.61$ km). Thus, for an elongated bottle resonator, the FSR of axial modes (determined by quantum number $q$) is much smaller than that of the azimuthal modes (determined by quantum number $m$).

We assume that the height of the bottle resonator is relatively small, $\Delta r_0 \ll r_0$ (Fig. 1(a)). Without losing the generality, we consider modes with $p = 0$, i.e., those most adjacent to the microresonator surface. The bandwidth $\Delta\Lambda_{ax}$ of series of axial eigenfrequencies supported by a bottle resonator with height $\Delta r_0$ is [13, 17] (Fig. 1(b)):

$$\Delta\Lambda_{ax} = \nu_0 \frac{\Delta r_0}{r_0} \quad (4)$$

where $\nu_0$ is the central frequency of the band. The total number of axial eigenfrequencies in this bandwidth is $\Delta\Lambda_{ax}/\Delta\nu_{ax}$. Depending on the height of the bottle resonator $\Delta r_0$, the bandwidth $\Delta\Lambda_{ax}$ can be smaller or larger than the azimuthal FSR $\Delta\nu_{az}$ (Fig. 1(b)-(d)).

For the effective generation of broadband and low repetition rate combs, the axial eigenfrequencies should match the azimuthal eigenfrequencies (i.e., $\Delta\nu_{az}$ should be a multiple of $\Delta\nu_{ax}$) as illustrated in Fig. 1(d). At the same time it is required that all the axial series are continuously spread along the broadband azimuthal series (Fig. 1(d)). To this end, the axial bandwidth $\Delta\Lambda_{ax}$ should be equal or exceed the azimuthal FSR $\Delta \nu_{az}$. From Eq. (3) and (4), the condition $\Delta \Lambda_{ax} \geq \Delta \nu_{az}$ is equivalent to

$$\Delta r_0 \geq \frac{c}{2\pi n_0 \nu_0} = \frac{\lambda_0}{2\pi n_0} \quad (5)$$

where $\lambda_0$ is the radiation frequency. For example, at $\lambda_0 \leq 2$ μm, the height of a silica bottle resonator can be as small as $\Delta r_0 \simeq 220$ nm. Assuming that the azimuthal FSR, $\Delta \nu_{az} \simeq c/(2\pi n_0 r_0)$, fits exactly $N$ axial FSRs, $\Delta \nu_{ax} = c(2\pi n_0)^{-1}(r_0 R_0)^{-1/2}$, we find:

$$\left(\frac{R_0}{r_0}\right)^{1/2} = N \quad (6)$$

Similar to the generation of OFCs at the series of azimuthal eigenfrequencies of silica WGM resonators, the negative dispersion, caused from the second term $\sim m^{1/3}$ in square brackets of Eq. (3), can be compensated at $\lambda_0 \sim 1.5$ μm by material dispersion [6-10]. In addition, the dispersion of azimuthal modes can be zeroed at an arbitrary radiation frequency (e.g., for visible light) if a silica fiber capillary rather than a solid fiber is used for the fabrication of the bottle resonator [18]. To this end, the thickness of the capillary can be appropriately reduced by internal etching with hydrofluoric acid, which can be continuously controlled optically [22, 23].

Remarkably, the dependence on the axial quantum number in Eq. (3) is exactly linear and therefore does not cause dispersion. A small correction to the shape of the bottle resonator may be necessary to compensate for the material dispersion and pump-detuning effect [6, 8], the question not addressed here.

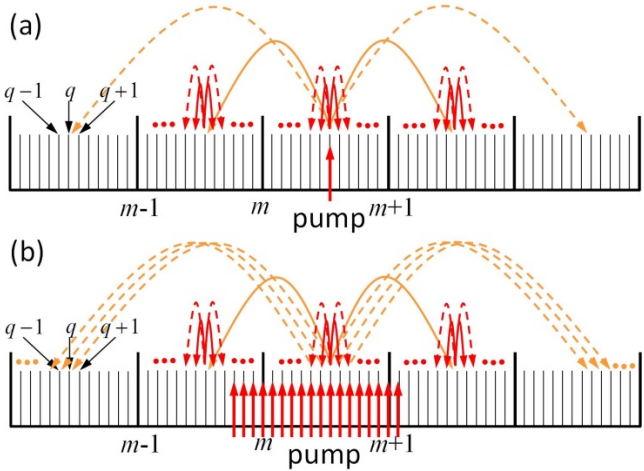

Fig 2. Illustration of the broadband and low RR OFCs generation in a bottle resonator with a single frequency cw laser pump (a) and with a narrowband comb pump generated by a mode-locked laser (b).

We propose two major approaches enabling the generation of broadband and low RR OFCs with the described microresonator. The first approach [11] consists in pumping the resonator with a cw laser at the radiation frequency corresponding to an eigenfrequency with quantum numbers $(m,q)$ (Fig. 2(a)). In this case, the comb frequencies corresponding to the azimuthal series $(m\pm 1,q)$, $(m\pm 2,q)$, $(m\pm 3,q)$, are generated following a nonlinear parametric excitation process similar to that experimentally demonstrated in toroidal resonators [6-8, 10]. Simultaneously, a frequency comb at equidistant axial frequencies $(m',q\pm 1)$, $(m',q\pm 2)$, $(m',q\pm 3)$, ... can be generated by means of the non-linear parametric excitation of axial modes. The latter process has been experimentally demonstrated in [9] and confirmed by recent numerical simulations [14]. Overall, a low RR broadband comb combining multiple series of low RR azimuthal peaks and the series of greater RR azimuthal peaks is generated.

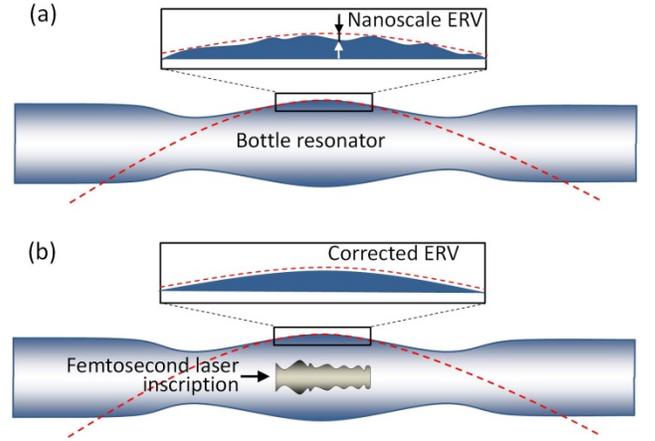

Fig. 3. Illustration of the ultraprecise fabrication process of a bottle resonator with characteristic ERV of several hundred of nanometers. (a) – Creation of a bottle microresonator by tapering an optical fiber. (b) – Correction of nanoscale deviations of the ERV from the parabolic dependence with a femtosecond laser inscription.

The second approach consists in employing a mode-locked laser generating a narrow-band low RR comb. It is assumed that the teeth of this comb match a series of $N$ axial eigenfrequencies of our resonator, $(m,q_1), (m,q_1+1),...,(m,q_1+N)$, situated along the bandwidth which is equal or greater than the azimuthal FSR (Fig. 2(b)). Pumping with this comb will generate similar series corresponding to different azimuthal quantum numbers $m, m\pm 1, m\pm 2,...$ and, consequently, a broadband and low RR comb identical to that of the first approach. The described comb generation is based on the process of nonlinear parametric excitation of azimuthal comb series employed in previous approaches [6-8]. This generation will be complemented by the parametric excitation within single axial series similar to that of the first approach. We conjecture that the mutual action of both processes (i.e., excitation within the azimuthal series and within the axial series) may enhance the nonlinear mode pulling process [5, 8] and further improve the equidistant character of the created broadband and low RR comb.

Fabrication of the proposed bottle microresonator frequency comb generator requires very accurate introduction of ERV, which,

from Eq. (5), should have the characteristic magnitude of a few hundreds of nanometers. Recently, a subangstrom precise fabrication of SNAP resonant structures and, in particular, bottle resonators with parabolic ERV, has been demonstrated [16, 21, 22]. The ultraprecise fabrication methods developed in SNAP are based on (i) the release of frozen-in stresses by annealing with a $CO_2$ laser [13] and (ii) the introduction of stresses by the femtosecond laser inscription [22]. A method expanding the developed technology to the fabrication of parabolic bottle resonators with the height of several hundred nanometers is illustrated in Fig. 3. First, softening and tapering a silica fiber allows to introduce a bottle resonator which ERV is close to parabolic in the vicinity of the resonator center and deviates from this dependence away from the center (Fig. 3(a)). In particular, nanometer-scale deviations from the parabolic dependence, caused by noise introduced during the tapering process [23], can be expected (inset in Fig. 3(a)). Further improvement of the fabrication precision can be achieved by post-processing with a femtosecond laser [22] (Fig. 3(b)). Specifically, after characterization of the ERV of fabricated bottle resonator [24, 13], the measured deviation can be corrected by iterations (inset in Fig. 3(b)). The correction process is similar to that described in [25].

In summary, we propose a bottle microresonator which can potentially serve as a broadband and low repetition rate optical frequency comb generator. The small bandwidth and small FSR axial eigenfrequency series of of this resonator are adjacent to the broadband and much greater FSR azimuthal series and consequently match each other. As the result, the resonator spectrum consists of small FSR, broadband, and equidistant eigenfrequencies. We suggest that pumping of this microresonator with a cw laser or, alternatively, with a small bandwidth and low RR comb of a mode-lock laser will allow us to generate an optical frequency comb, which is simultaneously low repetition rate and broadband. A method of ultraprecise fabrication of the designed microresonator is proposed.

**Funding.** Royal Society (WM130110). Horizon 2020 (H2020-EU.1.3.3, 691011).

**Acknowledgments**. MS acknowledges the Royal Society Wolfson Research Merit Award.